\documentclass[12pt,thmsa]{article}
\usepackage{amssymb}

\input{tcilatex}
\begin{document}

\title{Dark matter as an effect of the quantum vacuum}
\author{Emilio Santos \\
Departamento de F\'{i}sica\\
Universidad de Cantabria\\
Avenida de los Castros, s/n\\
Santander 39005. Spain\\
email: santose@unican.es}
\maketitle

\begin{abstract}
The interaction between the quantum vacuum and a weak gravitational field is
calculated for the vacuum fields of quantum electrodynamics. The result
shows that the vacuum state is modified by the gravitational field, giving
rise to a nonzero interaction energy. This suggests a model that fits in the
main properties of the hypothetical dark matter in galactic haloes.
\end{abstract}

\section{Introduction}

In quantum theory the vacuum is not empty but filled with quantum fields.
The fields have energy but in most applications of quantum mechanics that
energy may be ignored using the ``normal ordering rule'' that effectively
removes it. This fact has led some people to believe that the vacuum energy
is an artifact of the quantization procedure. However the belief is
difficult to reconcile with observable effects like the Lamb shift or the
Casimir effect\cite{Milonni}. An additional problem is that a simple attempt
to calculate the vacuum energy leads to infinities. The problem was shelved,
if not fully solved, with the renormalization techniques developed for
quantum electrodynamics (QED).

In applications of quantum mechanics not involving gravity the zero of
energy can be fixed arbitrarily, and putting it at the level of the vacuum
is the obvious choice. However when gravity is involved this is no longer
possible because energy gravitates. An attempt to solve the problem has been
the hypothesis that there is a natural cutoff making the vacuum energy (and
pressure) actually finite. Thus it is assumed that the vacuum stress-energy
tensor gives rise to a cosmological constant. However the apparently
plausible assumption that the cutoff should be obtained via combining the
universal constants $c,
\rlap{\protect\rule[1.1ex]{.325em}{.1ex}}h%
,G$ leads us to the Planck scale with a cosmological term too big by about
123 orders of magnitude. There is not yet a fully satisfactory solution to
the difficulty usually named the Cosmological Constant Problem\cite{Weinberg}%
. A simple solution is to assume that there is an effective cutoff, say at
the scale of a mass $m$. This solution was put forward by Zeldovich in 1976%
\cite{Zeldovich} that proposed for the energy density of the cosmological
term the following 
\begin{equation}
\rho _{DE}=\frac{Gm^{6}c^{2}}{
\rlap{\protect\rule[1.1ex]{.325em}{.1ex}}h%
^{4}}.  \label{Zel}
\end{equation}
This led Zeldovich to assume that the cosmological term derives from the
quantum vacuum fluctuations. The hypothesis has been revisited recently\cite
{Santos} as the origin of the ``dark energy'' needed to explain the
accelerated expansion of the universe\cite{expansion}. Indeed eq.$\left( \ref
{Zel}\right) $ fits in the current density of dark energy for $m\simeq
170Mev/c^{2},$ not much larger than the pion mass.

According to general relativity gravitation is curvature of spacetime,
consequently the problem of the infinities reappears in the study of quantum
fields in curved space. Indeed renormalization has been adapted to fields in
curved space and there exists a good perturbative approach to quantum
gravity in the framework of low energy effective field theory. A standard
reference on the subject is the book by Birrell and Davies\cite{Birrell},
but a number of review articles and many shorter introductory lecture notes
are also available (see e.g. \cite{Jacobson}) .

Following a general interpretation of quantum theory\cite{FOS}, in this
paper it is proposed that the vacuum is not an inert system, but its state
may be modified by the presence of matter, in particular by gravitational
interaction. It is also proposed that for weak gravitational fields such
interactions may be treated, avoiding the use of quantum gravity, via a
combination of quantum field theory and Newtonian gravity. One of these
cases is studied in section 2 and it suggests an alternative to (or
explanation for the effects of) dark matter. A specific model is presented
in section 3.

\section{Effect of a gravitational field on the quantum vacuum}

In this section I will study the interaction between the quantum vacuum and
a gravitational field. The subject is usually treated under the name quantum
fields in curved spacetime\cite{Birrell}. The most celebrated predictions
are Hawking radiation from black holes and the Unruh-Davis effect. Here I
study the action of a weak gravitational field on the vacuum with a method
that has some analogy with the calculation of the effect of an electrostatic
field, treated as classical, on the electron-positron vacuum field that
gives rise to vacuum polarization.

\subsection{The Newtonian approximation}

In the calculation I will use the Newtonian approximation. In Newtonian
gravity combined with special relativity the source of the field is any mass
or energy distribution and the interaction may be described via the product
of the mass (or energy) density, $\rho \left( \mathbf{r}\right) ,$ times the
Newtonian potential $\phi \left( \mathbf{r}\right) $. Of course Newtonian
gravity is not Lorentz invariant and therefore it is incompatible with
special relativity, but here I will study stationary systems where Lorentz
invariance plays no role. If we treat the gravitational field as classical
the quantum interaction Hamiltonian should be 
\begin{equation}
\hat{H}_{vac-grav}=\int d^{3}\mathbf{r}\hat{\rho}\left( \mathbf{r}\right)
\phi \left( \mathbf{r}\right) ,  \label{n1}
\end{equation}
where $\hat{\rho}\left( \mathbf{r}\right) $ is the operator for the energy
density of the quantum fields. This is similar to the QED treatment of the
electron-positron field in an external electromagnetic field, the latter
treated as classical. In this section I will use Planck units $
\rlap{\protect\rule[1.1ex]{.325em}{.1ex}}h%
=c=G=1,$ but I will write explicity Newton\'{}s constant $G$ in some cases
for the sake of clarity. Thus the the gravitational potential appears as
dimensionless.

The connection of eq.$\left( \ref{n1}\right) $ with a treatment in
semiclassical general relativity may be sketched as follows. Our aim would
be to find the energy of the ground state of a system consisting of quantum
fields in quantized spacetime. Now for the sake of clarity I consider a
single scalar field, $\chi ,$ with minimal coupling. Then classically the
dynamics of the system would derive from the action

\begin{equation}
S=\frac{1}{2}\int d^{4}x\sqrt{\left| g\right| }\left( g^{\mu \nu }\partial
_{\mu }\chi \partial _{\nu }\chi -m^{2}\chi ^{2}\right) .  \label{n1a}
\end{equation}
We should quantize both the field and the metric, but this would require
quantum gravity. The semiclassical theory may be developed as follows. Let
us assume that the ground state of the system is $\mid \Psi \rangle $ (that
we will determine later with some degree of accuracy). Then the
semiclassical approximation consists of calculating the metric from the
following Einstein tensor 
\begin{equation}
G_{\mu \nu }=\langle \Psi \left| \hat{T}_{\mu \nu }\left( x\right) \right|
\Psi \rangle .  \label{n2a}
\end{equation}
Putting that metric in eq.$\left( \ref{n1a}\right) $ it would be
straightforward to get the Hamiltonian and quantize the field, whence
getting the state $\mid \Psi \rangle $ would amount to solve a standard
eigenvalue problem.

The proposed method presents difficulties however. Firstly the quantity eq.$%
\left( \ref{n2a}\right) $ is infinite because it contains the product of
field operators at the same point. One way to make sense of the expectation
value is via the difference between the values in two different states. One
of these states would be the quantum vacuum in Minkowski space and the other
one may be a state with some mass distribution (e. g. corresponding to a
galaxy, see below). Another difficulty is that in order to get the metric we
need the quantum state $\mid \Psi \rangle $ of the system in eq.$\left( \ref
{n2a}\right) ,$ but in order to get the quantum state we need the metric in
eq.$\left( \ref{n1a}\right) $. The latter difficulty may be avoided in the
case of static weak fields where I may use the Newtonian approximation.

In the Newtonian approximation the metric may be approximated by 
\begin{equation}
g_{00}=1+2\phi \left( \mathbf{r}\right) ,g_{11}=g_{22}=g_{33}=-1,g_{\mu \nu
}=0\text{ if }\mu \neq \nu .  \label{n3a}
\end{equation}
where $\phi \left( \mathbf{r}\right) $ is the Newtonian potential derived
from the mass-energy present in the system via Newton's law, that now
replaces Einstein equation. Hence eq.$\left( \ref{n1a}\right) $ leads to 
\begin{eqnarray}
S &=&\frac{1}{2}\int dtd^{3}\mathbf{r}\sqrt{1+2\phi }\left[ \left( \frac{%
\partial \chi }{\partial t}\right) ^{2}-\left( \nabla \chi \right)
^{2}-m^{2}\chi ^{2}\right]  \nonumber \\
&\simeq &\frac{1}{2}\int dtd^{3}\mathbf{r}\left( 1+\phi \right) \left[
\left( \frac{\partial \chi }{\partial t}\right) ^{2}-\left( \nabla \chi
\right) ^{2}-m^{2}\chi ^{2}\right] ,  \label{n3b}
\end{eqnarray}
where $t$ is the time measured in the local Minkowski frame, that is 
\[
dt^{2}=g_{00}\left( dx^{0}\right) ^{2}, 
\]
and the second equality takes the weak field inequality $\left| \phi \right|
<<1$ into account. Hence it is straightforward to get the Hamiltonian for
this particular case, that I will not write. I will assume that a similar
result is valid in general, thus leading to eq.$\left( \ref{n1}\right) $. I
point out that we study a stationary system and the (approximate) metric eq.$%
\left( \ref{n3a}\right) $ corresponds to a flat space where the vector $%
\mathbf{r}$ makes sense.

We might believe that the vacuum is an inert stuff with nil stress-energy.
In this case the presence of a gravitational field is unable to change the
vacuum energy, which leads to the assumption that the interaction energy
between the field and the quantum vacuum should be given by 
\begin{equation}
E_{vac-grav}=-\int d^{3}\mathbf{r}d^{3}\mathbf{r}^{\prime }\frac{G\rho
_{vac}\left( \mathbf{r}\right) \rho \left( \mathbf{r}^{\prime }\right) }{%
\left| \mathbf{r-r}^{\prime }\right| }=\rho _{vac}\int d^{3}\mathbf{r}\phi
\left( \mathbf{r}\right) =0,  \label{2n}
\end{equation}
where 
\[
\rho _{vac}\equiv \langle \Psi _{vac}\left| \hat{\rho}\right| \Psi \rangle
_{vac} 
\]
is the vacuum expectation of the quantum energy density operator. This
density does not depend on the position vector, therefore it may be put
outside the space integral under the assumption that the vacuuum energy
density $\rho _{vac}$ is nil whence the interaction energy $E_{vac-grav}$ is
zero, as in the latter eq.$\left( \ref{2n}\right) $. Even if this hypothesis
does not hold true the former eq.$\left( \ref{2n}\right) $ means that the
vacuum is not influenced by the gravitational field created by matter$.$
However eq.$\left( \ref{2n}\right) $ is wrong, the correct one being eq.$%
\left( \ref{n1}\right) $ that takes into account the action of the field on
the quantum vacuum.

In the following I show that nonzero contributions exist to the
gravitational interaction between the vacuum fields and baryonic matter. In
section 3 I will study the proposal that this interaction may be the origin
of the socalled ``dark matter''. Here I sketch an explicit calculation of
some contributions to vacuum energy in quantum electrodynamics (QED),
firstly in the absence of gravity (subsection 2.2) and then with a
gravitational field present (subsection 2.3).

\subsection{The quantum vacuum of QED}

In this subsection I recall a few results of QED that, although well known,
will be used in later parts in the paper. Units $c= 
\rlap{\protect\rule[1.1ex]{.325em}{.1ex}}h%
=1$ will be used throughout so that $k$ may equally represent wavevector,
frequency or energy.

The properties of the vacuum electromagnetic field (commonly named zeropoint
field) are summarized as follows\cite{Milonni}. The spectrum is proportional
to the cube of the frequency and it is Lorentz invariant provided that a
cutoff is not placed (whence the energy density diverges quartically with
the frequency), but any cutoff breaks that symmetry.

The quantum operator of the energy density (or Hamiltonian density) is 
\begin{eqnarray}
\hat{\rho}_{EM} &\equiv &\frac{1}{2}\left( \mathbf{\hat{E}}^{2}+\mathbf{\hat{%
B}}^{2}\right) =\frac{1}{4V}\sum_{\mathbf{k,\varepsilon }}\sum_{\mathbf{k}%
^{\prime }\mathbf{,\varepsilon }^{\prime }}\{\sqrt{kk^{\prime }}[\mathbf{%
\varepsilon }\cdot \mathbf{\varepsilon }^{\prime }+\frac{1}{kk^{\prime }}%
\left( \mathbf{k\times \varepsilon }\right) \mathbf{\cdot }\left( \mathbf{k}%
^{\prime }\mathbf{\times \varepsilon }^{\prime }\right) ]  \nonumber \\
&&\times [\hat{\alpha}_{\mathbf{k},\mathbf{\varepsilon }}\exp (i\mathbf{k.r-}%
ikt)+\hat{\alpha}_{\mathbf{k},\mathbf{\varepsilon }}^{\dagger }\exp (-i%
\mathbf{k.r+}ikt)]  \nonumber \\
&&\times \left[ \hat{\alpha}_{\mathbf{k}^{\prime },\mathbf{\varepsilon }%
^{\prime }}\exp (i\mathbf{k}^{\prime }\mathbf{.r-}ik^{\prime }t)+\hat{\alpha}%
_{\mathbf{k}^{\prime },\mathbf{\varepsilon }^{\prime }}^{\dagger }\exp (-i%
\mathbf{k}^{\prime }\mathbf{.r+}ik^{\prime }t)\right] \},  \label{2.2c}
\end{eqnarray}
where $k\equiv \left| \mathbf{k}\right| ,k^{\prime }\equiv \left| \mathbf{k}%
^{\prime }\right| ,\mathbf{\varepsilon }$ is the polarization vector that
fulfils $\mathbf{\varepsilon \cdot k=0}$, and similar for $\mathbf{%
\varepsilon }^{\prime }.$ For later convenience we will rewrite eq.$\left( 
\ref{2.2c}\right) $ (ingoring the time dependence) in the form 
\begin{eqnarray}
\hat{\rho}_{EM} &=&\hat{\rho}_{EM1}+\hat{\rho}_{EM2},  \nonumber \\
\hat{\rho}_{EM1} &=&\frac{1}{2V}\sum_{\mathbf{k,\varepsilon }}\sum_{\mathbf{k%
}^{\prime }\mathbf{,\varepsilon }^{\prime }}\sqrt{kk^{\prime }}\mathbf{%
\varepsilon }*\mathbf{\varepsilon }^{\prime }\hat{\alpha}_{\mathbf{k}%
^{\prime },\mathbf{\varepsilon }^{\prime }}^{\dagger }\hat{\alpha}_{\mathbf{k%
},\mathbf{\varepsilon }}\exp \left[ i\left( \mathbf{k-k}^{\prime }\right)
\cdot \mathbf{r}\right] +\frac{1}{2V}\sum_{\mathbf{k,\varepsilon }}k, 
\nonumber \\
\hat{\rho}_{EM2} &=&\frac{1}{4V}\sum_{\mathbf{k,\varepsilon }}\sum_{\mathbf{k%
}^{\prime }\mathbf{,\varepsilon }^{\prime }}\sqrt{kk^{\prime }}\mathbf{%
\varepsilon }*\mathbf{\varepsilon }^{\prime }\hat{\alpha}_{\mathbf{k},%
\mathbf{\varepsilon }}\hat{\alpha}_{\mathbf{k}^{\prime },\mathbf{\varepsilon 
}^{\prime }}\exp \left[ i\left( \mathbf{k+k}^{\prime }\right) \cdot \mathbf{r%
}\right] +h.c.,  \label{2.2}
\end{eqnarray}
where \textit{h.c}. means Hermitean conjugate and for notational simplicity
I have labelled 
\begin{equation}
\mathbf{\varepsilon }*\mathbf{\varepsilon }^{\prime }\equiv \mathbf{%
\varepsilon }\cdot \mathbf{\varepsilon }^{\prime }+\frac{1}{kk^{\prime }}%
\left( \mathbf{k\times \varepsilon }\right) \mathbf{\cdot }\left( \mathbf{k}%
^{\prime }\mathbf{\times \varepsilon }^{\prime }\right) .  \label{1}
\end{equation}
In $\hat{\rho}_{EM1}$ I have written the operators in normal order, taking
the commutation relations into account.

The Hamiltonian is obtained by performing a space integral of the energy
density eq.$\left( \ref{2.2}\right) $, that is 
\begin{eqnarray}
\hat{H}_{EM} &=&\lim_{V\rightarrow \infty }\int_{V}\hat{\rho}_{EM}\left( 
\mathbf{r}\right) d^{3}\mathbf{r}=\lim_{V\rightarrow \infty }\int_{V}[\hat{%
\rho}_{EM1}\left( \mathbf{r}\right) +\hat{\rho}_{EM2}\left( \mathbf{r}%
\right) ]d^{3}\mathbf{r}  \nonumber \\
&=&\sum_{\mathbf{k,\varepsilon }}k(\hat{\alpha}_{\mathbf{k},\mathbf{%
\varepsilon }}^{\dagger }\hat{\alpha}_{\mathbf{k},\mathbf{\varepsilon }}+%
\frac{1}{2}).  \label{HEM}
\end{eqnarray}
The term $\hat{\rho}_{EM2}$ does not contribute to the Hamiltonian, and
therefore to the quantum expectation value of the vacuum energy, because the
space integral of the latter eq.$\left( \ref{2.2}\right) $ leads to $\mathbf{%
k}^{\prime }\mathbf{=-k}$, whence 
\begin{equation}
\int_{V}\hat{\rho}_{EM2}\left( \mathbf{r}\right) d^{3}\mathbf{r=}\frac{1}{4V}%
\sum_{\mathbf{k,\varepsilon ,\varepsilon }^{\prime }}k\hat{\alpha}_{\mathbf{k%
},\mathbf{\varepsilon }}\hat{\alpha}_{-\mathbf{k},\mathbf{\varepsilon }%
^{\prime }}\mathbf{\varepsilon }*\mathbf{\varepsilon }^{\prime }=0,
\label{1.b}
\end{equation}
because, as may be easily proved, 
\begin{equation}
\sum_{\varepsilon \mathbf{,\varepsilon }^{\prime }}\mathbf{\varepsilon }*%
\mathbf{\varepsilon }^{\prime }=\sum_{\varepsilon \mathbf{,\varepsilon }%
^{\prime }}[\mathbf{\varepsilon }\cdot \mathbf{\varepsilon }^{\prime }-\frac{%
1}{kk^{\prime }}\left( \mathbf{k\times \varepsilon }\right) \mathbf{\cdot }%
\left( \mathbf{k\times \varepsilon }^{\prime }\right) ]=0.  \label{1.a}
\end{equation}

For the free electromagnetic field the vacuum state, $\mid 0\rangle ,$ may
be defined as the state with the minimal energy amongst the eigenvectors of
the operator eq.$\left( \ref{HEM}\right) .$ It is a state with zero photons
and it has the properties 
\[
\alpha _{\mathbf{k},\mathbf{\varepsilon }}\mid 0\rangle =0,\langle 0\mid
\alpha _{\mathbf{k},\mathbf{\varepsilon }}^{\dagger }=0, 
\]
whence the vacuum expectation of the energy density is 
\begin{equation}
\left\langle 0\left| \hat{\rho}_{EM}\right| 0\right\rangle =\frac{1}{V}\sum_{%
\mathbf{k,\varepsilon }}\frac{1}{2}
\rlap{\protect\rule[1.1ex]{.325em}{.1ex}}h%
k=\frac{1}{V}\sum_{\mathbf{k}}
\rlap{\protect\rule[1.1ex]{.325em}{.1ex}}h%
k,  \label{2.4}
\end{equation}
where the latter equality derives from the two possible polarizations. In
the limit $V\rightarrow \infty $ eq.$\left( \ref{2.4}\right) $ leads to the
following result if a cutoff, $\Lambda ,$ is introduced in the photon
energies 
\begin{eqnarray}
\rho _{EM} &=&\frac{1}{V}\sum_{\mathbf{k}}k\rightarrow \int k\left( 2\pi
\right) ^{-3}d^{3}k,  \nonumber \\
&=&\frac{1}{2\pi ^{2}}\int_{0}^{k_{\max }}k^{3}dk=\frac{\Lambda ^{4}}{8\pi
^{2}}.  \label{0}
\end{eqnarray}

For the electron-positron field the energy density may be written 
\begin{equation}
\hat{\rho}_{D}=\frac{i}{2}\left( \hat{\psi}^{\dagger }\frac{d\hat{\psi}}{dt}-%
\frac{d\hat{\psi}^{\dagger }}{dt}\hat{\psi}\right) ,  \label{2.5b}
\end{equation}
(the subindex D stands for Dirac). Expanding $\hat{\psi}$ and $\hat{\psi}%
^{\dagger }$ in plane waves we get, after some algebra, the following energy
density 
\begin{equation}
\hat{\rho}_{D}\left( \mathbf{r},t\right) =\rho _{D0}+\hat{\rho}_{b}\left( 
\mathbf{r},t\right) +\hat{\rho}_{d}\left( \mathbf{r},t\right) ,  \label{2.6}
\end{equation}
where 
\begin{eqnarray*}
\rho _{D0} &=&-\frac{1}{V}\sum_{p\mathbf{,}s}\sqrt{m^{2}+p^{2}}, \\
\hat{\rho}_{b}\left( \mathbf{r},t\right) &=&\frac{1}{V}\sum_{\mathbf{pp}%
^{\prime }ss^{\prime }}B\left( \mathbf{p,}s,\mathbf{p}^{\prime },s^{\prime
}\right) \hat{b}_{\mathbf{p}^{\prime }s^{\prime }}^{\dagger }\hat{b}_{%
\mathbf{p},s}\exp \left[ i\left( \mathbf{p-p}^{\prime }\right) \cdot \mathbf{%
r-}i\left( E-E^{\prime }\right) t\right] , \\
\hat{\rho}_{d}\left( \mathbf{r},t\right) &=&\frac{1}{V}\sum_{\mathbf{pp}%
^{\prime }ss^{\prime }}D\left( \mathbf{p,}s,\mathbf{p}^{\prime },s^{\prime
}\right) d_{\mathbf{p}^{\prime }s^{\prime }}^{\dagger }\hat{d}_{\mathbf{p}%
,s}\exp \left[ i\left( \mathbf{p-p}^{\prime }\right) \cdot \mathbf{r-}%
i\left( E-E^{\prime }\right) t\right] ,
\end{eqnarray*}
$\hat{b}_{\mathbf{p},s}$ $\left( \hat{d}_{\mathbf{p},s}\right) $ being the
annihilation operator of an electron (positron) with momentum $\mathbf{p}$
and spin $s(=1,2)$, $\hat{b}_{\mathbf{p},s}^{\dagger }\left( \hat{d}_{%
\mathbf{p},s}^{\dagger }\right) $ the corresponding creation operator. The
first term in eq.$\left( \ref{2.6}\right) $, $\rho _{D0},$ is a c-number,
not an operator (more properly it is proportional to the unit operator). Its
negative sign is a consequence of the anticommutation rules of fermi field
operators. Getting the functions $B$ and $D$\ is straightforward but I shall
not write them. Integration of eq.$\left( \ref{2.6}\right) $ with respect to 
$\mathbf{r}$\textbf{\ }gives the Hamiltonian of the free electron-positron
field, that is

\begin{equation}
\hat{H}_{D}=\sum_{p\mathbf{,}s}\sqrt{m^{2}+p^{2}}\left( \hat{b}_{\mathbf{p}%
,s}^{\dagger }\hat{b}_{\mathbf{p},s}+\hat{d}_{\mathbf{p},s}^{\dagger }\hat{d}%
_{\mathbf{p},s}-1\right) .  \label{HD}
\end{equation}

The vacuum state, $\mid 0\rangle ,$ of the free Dirac field corresponds to
the eigenvector of the Hamiltonian eq.$\left( \ref{HD}\right) $ with the
smallest eigenvalue. Therefore from now on we define $\mid 0\rangle $ to be
the QED free-fields vacuum state, which is a simultaneous eigenvector of
both Hamiltonians eqs.$\left( \ref{HEM}\right) $ and $\left( \ref{HD}\right)
,$ and consequently an eigenvector of the total free fields Hamiltonian,
that is 
\begin{equation}
\hat{H}_{0}\mid 0\rangle =\left( \hat{H}_{EM}+\hat{H}_{D}\right) \mid
0\rangle =E_{0}\mid 0\rangle ,  \label{H0}
\end{equation}
The state $\mid 0\rangle $ is defined as having zero photons, electrons and
positrons. It should be distinguished from the physical vacuum state, $\mid
vac\rangle ,$ which is an eigenvalue of the total Hamiltonian, including the
interactions.

The energy density of the free electron-positron field follows easily from
eq.$\left( \ref{HD}\right) $. It is \textit{negative} and divergent.
Introducing an energy cut-off $\Lambda $ for the electrons and positrons, we
get 
\begin{eqnarray}
\rho _{D} &=&-\frac{1}{V}\sum_{p\mathbf{,}s}\sqrt{m^{2}+p^{2}}\rightarrow
-\pi ^{-2}\int_{0}^{p_{\max }}\sqrt{m^{2}+p^{2}}p^{2}dp=-\pi
^{-2}\int_{m}^{\Lambda }\sqrt{E^{2}-m^{2}}E^{2}dE  \nonumber \\
&=&-\frac{1}{4\pi ^{2}}\left[ \Lambda (\Lambda ^{2}-\frac{1}{2}m^{2})\sqrt{%
\Lambda ^{2}-m^{2}}-\frac{1}{2}m^{4}\cosh ^{-1}\left( \frac{\Lambda }{m}%
\right) \right]  \nonumber \\
&=&-\frac{1}{4\pi ^{2}}\left[ \Lambda ^{4}-\Lambda ^{2}m^{2}+\frac{1}{8}%
m^{4}-\frac{1}{2}m^{4}\ln \left( \frac{2\Lambda }{m}\right) \right] +O\left(
\Lambda ^{-2}\right) ,\smallskip \smallskip  \label{5}
\end{eqnarray}
The negative value might be anticipated by inspection of the Hamiltonian eq.$%
\left( \ref{HD}\right) .$ The total free fields vacuum energy is the product
of the integration volume, $V$, times the energy density, $\rho _{0}=\rho
_{EM}+\rho _{D}.$

The physical vacuum of QED, $\mid vac\rangle ,$\ is different from the free
field vacuum, $\mid 0\rangle ,$\ studied above. The latter is the
eigenvector, with the smallest eigenvalue, of the Hamiltonian $H_{0},$\ see
eq.$\left( \ref{H0}\right) .$\ The former is an eigenvector of the total
Hamiltonian $H=H_{0}+H_{int}$ that takes the interaction into account.
Finding $\mid vac\rangle $\ as an exact eigenvector of $H$\ is not possible
in practice and it is standard to use a perturbation method, that leads to 
\begin{equation}
\mid vac\rangle =c_{0}\mid 0\rangle +\sum_{n=0}c_{n}\mid n\rangle ,
\label{vac}
\end{equation}
the states $\mid 0\rangle $ and $\left\{ \mid n\rangle \right\} $ being
eigenstates of the unperturbed Hamiltonian $H_{0}$. Usually the sum eq.$%
\left( \ref{vac}\right) $ gives an expansion in powers of the coupling
constant, the electron charge $e$. Only even powers of $e$\ would appear and
the result becomes an expansion in powers of the fine structure constant $%
\alpha \equiv e^{2}/(4\pi 
\rlap{\protect\rule[1.1ex]{.325em}{.1ex}}h%
c)\simeq 1/137.$ The perturbation method is sensible if 
\begin{equation}
\sum_{n}\left| c_{n}\right| ^{2}=1,\left| c_{n\neq 0}\right| ^{2}<<1.
\label{vac1}
\end{equation}
Actually the former condition is not fulfilled if we take $c_{0}=1$ as
usual, but a normalization is possible if the sum converges, that is 
\begin{equation}
\sum_{n}\left| c_{n}\right| ^{2}<\infty .  \label{vac2}
\end{equation}

The Hamiltonian (or energy) density operator for the interaction in QED may
be written, in the Coulomb gauge, 
\begin{equation}
\hat{\rho}_{int}\left( \mathbf{r,}t\right) =-e\hat{\psi}^{\dagger }\mathbf{%
\alpha }\hat{\psi}\cdot \mathbf{\hat{A}.}  \label{ront}
\end{equation}
The operators $\hat{\psi},\hat{\psi}^{\dagger }$\ and $\mathbf{\hat{A}}$\
contain two terms each when expanded in plane waves, every term
corresponding to an infinite sum. One of these terms has creation operators
and the other one annihilation operators. This gives rise to 8 terms for $%
\hat{\rho}_{int},$\ eq.$\left( \ref{2.2c}\right) $. I will write only the
two terms that will survive in the Hamiltonian. We get (ignoring the time
dependence that is irrelevant in the following) 
\begin{eqnarray}
\hat{\rho}_{int}\left( \mathbf{r}\right) &=&\sum_{\mathbf{p,q},\mathbf{k}%
,s,s^{\prime },\varepsilon }\left[ \zeta _{n}\hat{\alpha}_{\mathbf{k},%
\mathbf{\varepsilon }}\hat{b}_{\mathbf{p}s}\hat{d}_{\mathbf{q}s^{\prime
}}\exp \left[ i\left( \mathbf{p+q+k}\right) \cdot \mathbf{r}\right]
+h.c\right]  \nonumber \\
\zeta _{n} &\equiv &-e\frac{m}{V^{1/2}\sqrt{2kEE^{\prime }}}u_{s}^{\dagger
}\left( \mathbf{p}\right) \mathbf{\alpha \cdot \varepsilon }v_{s^{\prime
}}\left( \mathbf{q}\right) ,  \label{Hint}
\end{eqnarray}
where $h.c.$\ means Hermitean conjugate, $u_{s}^{\dagger }$\textbf{\ }and $%
v_{s^{\prime }}$ are spinors, 
\[
E\equiv \sqrt{p^{2}+m^{2}},E^{\prime }\equiv \sqrt{q^{2}+m^{2}}, 
\]
and $n$ stands for $\left\{ \mathbf{p,q},\mathbf{k},s,s^{\prime
},\varepsilon \right\} .$ The interaction Hamiltonian $\hat{H}_{int}$ is the
space integral of $\hat{\rho}_{int}\left( \mathbf{r}\right) $ within the
volume $V$. One of the terms of the Hamiltonian may create triples
electron-positron-photon and the other term may annihilate triples.

It is easy to get the matrix element between the vacuum and a state with one
triple $e^{-}e^{+}\gamma $. The Hamiltonian is obtained via a space integral
and the interaction energy may be calculated to second order perturbation
theory giving (tentatively, see below) 
\begin{eqnarray}
E_{int} &=&V\rho _{int}=-\sum_{n}\frac{\left| \left\langle 0\left| \hat{H}%
_{int}\right| n\right\rangle \right| ^{2}}{k+E+E^{\prime }}  \label{21.5} \\
&=&-\sum_{\mathbf{p,q,k}}\frac{\sum_{s,s^{\prime },\varepsilon }\left| \zeta
_{n}\right| ^{2}}{k+E+E^{\prime }}\left| \int_{V}\exp \left[ i\left( \mathbf{%
p+q+k}\right) \cdot \mathbf{r}\right] d^{3}\mathbf{r}\right| ^{2}.  \nonumber
\end{eqnarray}
Taking the definition of $\zeta _{n},$ eq.$\left( \ref{Hint}\right) $ into
account we get, after some algebra, 
\begin{equation}
E_{int}=-\frac{e^{2}}{V^{3}}\sum_{\mathbf{p,q,k}}\frac{\left[
(p^{2}+q^{2})k^{2}+(\mathbf{p}\cdot \mathbf{k)(q}\cdot \mathbf{k)}\right] }{%
2k^{3}EE^{\prime }\left( k+E+E^{\prime }\right) }\left| \int_{V}\exp \left[
i\left( \mathbf{p+q+k}\right) \cdot \mathbf{r}\right] d^{3}\mathbf{r}\right|
^{2}.  \label{21.6}
\end{equation}
The integrals of $\mathbf{r}$\ are trivial and give the square of the
integration volume, $V,$ times a Kroneker delta $\delta _{\mathbf{p+q,-k}}$
of momentum conservation. The sums in $\mathbf{p}$ and $\mathbf{q}$ give
rise to a quartic divergence similar to those found for the free fields. To
make easier the calculation we may introduce an ultraviolet cutoff via the
multiplication of eq.$\left( \ref{21.6}\right) $ times the convergence
factor 
\begin{equation}
\exp \left[ -\gamma \left( \left( k+E+E^{\prime }\right) \right) \right] ,
\label{cutoff}
\end{equation}
where $\varepsilon $ is of order the inverse of the parameter $\Lambda $ of
eqs.$\left( \ref{0}\right) $ and $\left( \ref{5}\right) .$ We are interested
in the energy density $E_{int}/V,$ see eq.$\left( \ref{21.6}\right) ,$ that
after including the convergence factor eq.$\left( \ref{cutoff}\right) ,$ I
shall label $A$. The calculation might be performed expanding $E=\sqrt{%
p^{2}+m^{2}}$ in powers of $m^{2}$ and similar for $E^{\prime }$ but it
would be lengthy. It is not reported here but it may be realized that it
should negative and proportional to $\gamma ^{-4}$, that is having the form 
\begin{equation}
E_{int}\equiv A=-3A_{0}\gamma ^{-4},A_{0}>0,  \label{A}
\end{equation}
in the limit $m\rightarrow 0,$ that is a good approximation because $\gamma
^{-1}>>m.$ Here $A_{0}$ is a numerical parameter and the factor 3 is
introduced for later convenience.

Actually the result eq.$\left( \ref{A}\right) $ is not reliable because the
normalization condition, former eq.$\left( \ref{vac1}\right) ,$ is not
fulfilled. This problem may be solved dividing eq.$\left( \ref{21.6}\right) $
by the square root of the quantity $1+\sum_{n\neq 0}\left| c_{n}\right|
^{2}. $ In fact as is well known the second order perturbation for the
energy, eq.$\left( \ref{21.6}\right) ,$ may be written in terms of the first
order perturbation of the normalized statevector as follows 
\begin{equation}
E_{int}=V\rho _{int}=\sum_{n\neq 0}\frac{c_{n}}{\sqrt{1+\sum_{n\neq 0}\left|
c_{n}\right| ^{2}}}\left\langle 0\left| \hat{H}_{int}\right| n\right\rangle =%
\frac{A}{\sqrt{C}}.  \label{22}
\end{equation}
Here we define 
\begin{equation}
C\equiv 1+\sum_{n\neq 0}\left| c_{n}\right| ^{2}\simeq \sum_{n\neq 0}\left|
c_{n}\right| ^{2}=\sum_{n\neq 0}\frac{\left| \left\langle 0\left| \hat{H}%
_{int}\right| n\right\rangle \right| ^{2}}{(k+E+E^{\prime })^{2}}\exp \left[
-\varepsilon \left( \left( k+E+E^{\prime }\right) \right) \right] , 
\nonumber
\end{equation}
where I have included the regularization eq.$\left( \ref{cutoff}\right) .$
The approximate equality takes into account that $\sum_{n\neq 0}\left|
c_{n}\right| ^{2}>>1$. Thus we get 
\begin{equation}
A=dC/d\varepsilon \Rightarrow C=A_{0}\gamma ^{-3}.  \label{C}
\end{equation}
consistent with eq.$\left( \ref{A}\right) $ because, with the regularization
eq.$\left( \ref{cutoff}\right) $ included, theformer eq.$\left( \ref{C}%
\right) $ holds true as may be realized. Eq.$\left( \ref{22}\right) $
provides a better approximation than eq.$\left( \ref{A}\right) .$

As a conclusion of this subsection we exhibit the total QED vacuum energy
within a finite volume $V$ in the presence of a gravitational potential $%
\phi \left( \mathbf{r}\right) ,$ calculated as in eq.$\left( \ref{2n}\right)
.$ It is the following

\begin{equation}
E_{vac}^{QED}=\rho _{vac}\left[ V+\int d^{3}\mathbf{r}\phi \left( \mathbf{r}%
\right) \right] =\left( \rho _{EM}+\rho _{D}+\rho _{int}\right) \left[ V+%
\bar{\phi}\right] ,\bar{\phi}\equiv \int d^{3}\mathbf{r}\phi \left( \mathbf{r%
}\right) .  \label{n10}
\end{equation}
where we assume that the field $\phi \left( \mathbf{r}\right) $ is zero (or
negligible) outside the volume $V$. The result, a straightforward
consequence of eq.$\left( \ref{2n}\right) $, follows from the hypothesis
that the quantum vacuum is an inert system in the sense that its state is
not modified by a gravitational field. In the following I prove that the
result is quite different if we do not support that hypothesis but use eq.$%
\left( \ref{n1}\right) $.

\subsection{Interaction between the vacuum and a gravitational field in QED}

The aim of this subsection is to prove that, according eq.$\left( \ref{n1}%
\right) $ and at a difference with eq.$\left( \ref{2n}\right) ,$ a
gravitational field does modify the quantum vacuum state giving rise to a
nonzero interaction energy between the vacuum fields and the gravitational
field. For the proof I start rewriting both eqs.$\left( \ref{2n}\right) $
and $\left( \ref{n1}\right) $ separating in each the QED contribution from
all other contributions, that is 
\begin{eqnarray}
E_{inertvac-grav}^{total} &=&\left[ \rho _{vac}^{QED}+\rho
_{vac}^{other}\right] \int d^{3}\mathbf{r}\phi \left( \mathbf{r}\right) , 
\nonumber \\
\hat{H}_{vac-grav} &=&\int d^{3}\mathbf{r}\left[ \hat{\rho}^{QED}\left( 
\mathbf{r}\right) +\hat{\rho}^{other}\left( \mathbf{r}\right) \right] \phi
\left( \mathbf{r}\right) ,  \label{n}
\end{eqnarray}
where $\rho _{vac}^{QED}$ is the quantity eq.$\left( \ref{n10}\right) .$

The total quantum vacuum energy, $\rho _{vac}^{total}=\rho _{vac}^{QED}+\rho
_{vac}^{other},$ is known to be of order $\Lambda ^{4}$ where $\Lambda $ is
an energy (or inverse length) introduced as a cutoff, usually assumed to be
at the Planck scale. With this assumption the vacuum energy density is huge
but nevertheless it is possible to make calculations ignoring it (e. g. the
Lamb shift or the anomalous magnetic moment of the electron). Therefore some
mechanism would produce an effective cancellation that allows taking the
vacuum energy as nil or small in practice. The mechanism is not yet known
and the difficulty gives rise to the socalled Cosmological Constant Problem
mentioned in the introduction. In this paper I shall ignore that problem via
the simple expedience of assuming that the total vacuum energy is zero
(notice that for instance in QED there are positive contributions like eqs.$%
\left( \ref{5}\right) $ and $\left( \ref{22}\right) $ and negative ones like
eq.$\left( \ref{0}\right) ).$ That is I will assume 
\begin{equation}
\rho _{vac}^{total}=0\Rightarrow \rho _{vac}^{other}=-\rho _{vac}^{QED},
\label{n30}
\end{equation}
whence from eq.$\left( \ref{n}\right) $ we get $E_{inertvac-grav}^{total}=0.$

Now I will compare the interaction energy calculated from the Hamiltonian
latter eq.$\left( \ref{n}\right) ,$ that is 
\begin{equation}
\hat{H}_{vac-grav}^{QED}=\int d^{3}\mathbf{r}\hat{\rho}_{vac}^{QED}\left( 
\mathbf{r}\right) \phi \left( \mathbf{r}\right) ,  \label{n31}
\end{equation}
with the interaction energy obtained via the hypothesis involved in the
former eq.$\left( \ref{n}\right) ,$ that is 
\begin{equation}
E_{inertvac-grav}^{QED}=\rho _{vac}^{QED}\int d^{3}\mathbf{r}\phi \left( 
\mathbf{r}\right) .  \label{n32}
\end{equation}
The relevant result is that subtracting the energy calculated via eq.$\left( 
\ref{n31}\right) $ minus the energy eq.$\left( \ref{n32}\right) $ we will
obtain a nonzero quantity. Our hypothesis is that similar calculations with
all other (interacting) vacuum fields may give also a nonzero contribution
to the difference. I believe that this contribution is relevant in
astrophysics and proceed to calculating it for QED in the following.

I shall begin with the free electromagnetic field. In contrast with eq.$%
\left( \ref{n32}\right) ,$ finding the (approximate) eigenvalue of the
Hamiltonian eq.$\left( \ref{n31}\right) $ is made as follows. The
Hamiltonian density operator of the vacuum interacting with an external
gravitational field $\phi \left( \mathbf{r}\right) $ is 
\begin{equation}
\hat{\rho}_{vac-grav}^{_{EM}}=\left[ \hat{\rho}_{EM1}\left( \mathbf{r}%
\right) +\hat{\rho}_{EM2}\left( \mathbf{r}\right) \right] \left[ 1+\phi
\left( \mathbf{r}\right) \right] ,  \label{n33}
\end{equation}
whence the Hamiltonian is obtained performing a space integration. Taking
into account eq.$\left( \ref{2.2}\right) $ we get 
\begin{eqnarray}
\hat{H}_{vac-grav}^{_{EM}} &=&\hat{H}_{vac-grav}^{_{EM1}}+\hat{H}%
_{vac-grav}^{_{EM2}},  \label{3} \\
\hat{H}_{vac-grav}^{_{EM1}} &=&\frac{1}{2}\left[ 1+\frac{1}{V}\bar{\phi}%
\right] \left[ \sum_{\mathbf{k,\varepsilon }}k+\sum_{\mathbf{k,\varepsilon }%
}k\hat{\alpha}_{\mathbf{k},\mathbf{\varepsilon }}^{\dagger }\hat{\alpha}_{%
\mathbf{k},\mathbf{\varepsilon }}\right]  \nonumber \\
&&+\frac{1}{2V}\sum_{\mathbf{k,\varepsilon }}\sum_{\mathbf{k}^{\prime }%
\mathbf{,\varepsilon }^{\prime }}\sqrt{kk^{\prime }}\mathbf{\varepsilon }*%
\mathbf{\varepsilon }^{\prime }\hat{\alpha}_{\mathbf{k}^{\prime },\mathbf{%
\varepsilon }^{\prime }}^{\dagger }\hat{\alpha}_{\mathbf{k},\mathbf{%
\varepsilon }}\int \phi \left( \mathbf{r}\right) \exp \left[ i\left( \mathbf{%
k-k}^{\prime }\right) \cdot \mathbf{r}\right] d^{3}\mathbf{r,}  \nonumber \\
\hat{H}_{vac-grav}^{_{EM2}} &=&\frac{1}{4V}\sum_{\mathbf{k,\varepsilon }%
}\sum_{\mathbf{k}^{\prime }\mathbf{,\varepsilon }^{\prime }}\sqrt{kk^{\prime
}}\mathbf{\varepsilon }*\mathbf{\varepsilon }^{\prime }\hat{\alpha}_{\mathbf{%
k},\mathbf{\varepsilon }}\hat{\alpha}_{\mathbf{k}^{\prime },\mathbf{%
\varepsilon }^{\prime }}\int \phi \left( \mathbf{r}\right) \exp \left[
i\left( \mathbf{k+k}^{\prime }\right) \cdot \mathbf{r}\right] +h.c.. 
\nonumber
\end{eqnarray}
In comparison with the Hamiltonian obtained in absence of gravitational
field, eq.$\left( \ref{HEM}\right) ,$ we see that there are additional terms.

Now we shall find the \textit{EM} contribution to the vacuum energy via
solving the eigenvalue problem for the Hamiltonian $\hat{H}%
_{vac-grav}^{_{EM}}.$ Firstly it is easy to see that the zero photon state $%
\mid 0\rangle $ is the ground state for the Hamiltonian $\hat{H}%
_{vac-grav}^{_{EM1}}$ with eigenvalue (compare with eq.$\left( \ref{n10}%
\right) )$%
\begin{equation}
E_{vac-grav}^{_{EM1}}=\rho _{EM}\left[ V+\bar{\phi}\right] .  \label{3a}
\end{equation}
In fact the two latter terms of $\hat{H}_{vac-grav}^{_{EM1}}$ give zero when
acting on the state $\mid 0\rangle $ because they have an annihilation
operator in the right.

However 
\[
\hat{H}_{vac-grav}^{_{EM2}}\mid 0\rangle \neq 0,
\]
which shows that $\mid 0\rangle $ is no longer the quantum vacuum state of
the electromagnetic field. We conclude that \textit{the presence of a
gravitational field modifies the quantum vacuum state. }The modification
consists of the vacuum state becoming 
\[
\mid vac\rangle =c_{0}\mid 0\rangle +\sum_{n}c_{n}\mid n\rangle ,
\]
where $\left\{ \mid n\rangle \right\} $ is a set of quantum states with even
number of photons, in particular 2 photons if the coefficients $\left\{
c_{n}\right\} $ are calculated to first order in the perturbation $\hat{H}%
_{vac-grav}^{_{EM2}},$ eq.$\left( \ref{3}\right) .$ I will not get the
coefficients $\left\{ c_{n}\right\} $ but only the energy, this to second
order. We have 
\begin{eqnarray}
E_{vac-grav}^{_{EM2}} &=&\sum_{n}\frac{\left| \langle 0\left| \hat{H}%
_{vac-grav}^{_{EM2}}\right| n\rangle \right| ^{2}}{E_{0}-E_{n}}\smallskip  
\nonumber \\
&=&-\langle 0\mid \int d^{3}\mathbf{r}_{1}\hat{\rho}_{EM2}\left( \mathbf{r}%
_{1}\right) \phi \left( \mathbf{r}_{1}\right) \int d^{3}\mathbf{r}_{2}\hat{%
\rho}_{EM2}\left( \mathbf{r}_{2}\right) \phi \left( \mathbf{r}_{2}\right)
\mid 0\rangle /\left( k+k^{\prime }\right)   \nonumber \\
&=&-\frac{1}{V^{2}}\sum_{\mathbf{k,k}^{\prime }}\frac{kk^{\prime }}{%
k+k^{\prime }}\left[ 1+\frac{\mathbf{k\cdot k}^{\prime }}{kk^{\prime }}%
\right] ^{2}  \nonumber \\
&&\times \int d^{3}\mathbf{r}_{1}\int d^{3}\mathbf{r}_{2}\{\exp \left[
i\left( \mathbf{k}+\mathbf{k}^{\prime }\right) \mathbf{.(r}_{2}-\mathbf{r}%
_{1})\right] \phi \left( \mathbf{r}_{1}\right) \phi \left( \mathbf{r}%
_{2}\right) \},  \label{6}
\end{eqnarray}
where I have taken the latter eq.$\left( \ref{3}\right) $ into account.

After the change of variables 
\begin{equation}
\mathbf{r}_{1}=\mathbf{r}_{0}-\mathbf{r}/2,\mathbf{r}_{2}=\mathbf{r}_{0}+%
\mathbf{r}/2,  \label{6a}
\end{equation}
we get 
\begin{eqnarray}
E_{vac-grav}^{_{EM2}} &=&-\frac{1}{V^{2}}\sum_{\mathbf{k,k}^{\prime }}\frac{%
(kk^{\prime }+\mathbf{k\cdot k}^{\prime })^{2}}{(k+k^{\prime })kk^{\prime }}
\nonumber \\
&&\times \int d^{3}\mathbf{r}_{0}\int d^{3}\mathbf{r}\exp \left[ i\left( 
\mathbf{k}+\mathbf{k}^{\prime }\right) \mathbf{.r}\right] \phi \left( 
\mathbf{r}_{0}+\mathbf{r}/2\right) \phi \left( \mathbf{r}_{0}-\mathbf{r}%
/2\right) .  \label{2}
\end{eqnarray}
We may approximate 
\[
\phi \left( \mathbf{r}_{0}\pm \mathbf{r}/2\right) \simeq \phi \left( \mathbf{%
r}_{0}\right) \pm \frac{1}{2}\sum_{i}x_{i}\frac{\partial }{\partial x_{0i}}%
\phi \left( \mathbf{r}_{0}\right) +\frac{1}{8}\sum_{ij}x_{i}x_{j}\frac{%
\partial ^{2}}{\partial x_{0i}\partial x_{0j}}\phi \left( \mathbf{r}%
_{0}\right) ,
\]
whence, retaining only the relevant terms we may write, up to second order
in the components $x_{i},$%
\begin{eqnarray}
\phi \left( \mathbf{r}_{0}+\mathbf{r}/2\right) \phi \left( \mathbf{r}_{0}-%
\mathbf{r}/2\right)  &\simeq &\phi \left( \mathbf{r}_{0}\right) ^{2} 
\nonumber \\
&&+\frac{1}{12}r^{2}\left[ \phi \left( \mathbf{r}_{0}\right) \mathbf{\nabla }%
^{2}\phi \left( \mathbf{r}_{0}\right) -\left| \mathbf{\nabla }\phi \left( 
\mathbf{r}_{0}\right) \right| ^{2}\right] ,  \label{70}
\end{eqnarray}
As the integrals in $\mathbf{k,k}^{\prime }$ involve all directions of space
we have substituted $\frac{1}{3}r^{2}\left| \mathbf{\nabla }\phi \left( 
\mathbf{r}_{0}\right) \right| ^{2}$ for $\left[ \mathbf{r\cdot \nabla }\phi
\left( \mathbf{r}_{0}\right) \right] ^{2}$ and a similar change in the term
with $\mathbf{\nabla }^{2}\phi \left( \mathbf{r}_{0}\right) .$ I point out
that the latter term of eq.$\left( \ref{70}\right) $ is negative because $%
\phi <0,\mathbf{\nabla }^{2}\phi >0$. Therefore I will use the opposite to
the term in the following.

Inserting eq.$\left( \ref{70}\right) $ in eq.$\left( \ref{2}\right) $ it is
easy to see that the term with $\phi \left( \mathbf{r}_{0}\right) ^{2}$ is
nil because the space\textbf{\ }integral leads to $\mathbf{k}^{\prime }=-%
\mathbf{k}$, whence $kk^{\prime }+\mathbf{k\cdot k}^{\prime }=0.$ Then,
after taking the continuous limit of the sums in $\mathbf{k,k}^{\prime },$
we get for the energy 
\begin{eqnarray}
E_{vac-grav}^{_{EM2}} &=&-\frac{1}{12\left( 8\pi ^{3}\right) ^{2}}\int d^{3}%
\mathbf{k}\int d^{3}\mathbf{k}^{\prime }\frac{(kk^{\prime }+\mathbf{k\cdot k}%
^{\prime })^{2}}{(k+k^{\prime })kk^{\prime }}\int \mathbf{r}^{2}d^{3}\mathbf{%
r}\exp \left[ i\left( \mathbf{k}+\mathbf{k}^{\prime }\right) \mathbf{.r}%
\right]   \nonumber \\
&&\times \int d^{3}\mathbf{r}_{0}\left\{ -\left[ \left| \mathbf{\nabla }\phi
\left( \mathbf{r}_{0}\right) \right| ^{2}-\phi \left( \mathbf{r}_{0}\right) 
\mathbf{\nabla }^{2}\phi \left( \mathbf{r}_{0}\right) \right] \right\} , 
\nonumber \\
&=&\int d^{3}\mathbf{r}_{0}\left[ \left| \mathbf{\nabla }\phi \left( \mathbf{%
r}_{0}\right) \right| ^{2}-\phi \left( \mathbf{r}_{0}\right) \mathbf{\nabla }%
^{2}\phi \left( \mathbf{r}_{0}\right) \right] K_{EM},  \label{8}
\end{eqnarray}
where 
\[
K_{EM}\equiv \frac{1}{12\left( 8\pi ^{3}\right) ^{2}}\int d^{3}\mathbf{k}%
\int d^{3}\mathbf{k}^{\prime }\frac{(kk^{\prime }+\mathbf{k\cdot k}^{\prime
})^{2}}{(k+k^{\prime })kk^{\prime }}\int r^{2}d^{3}\mathbf{r}\exp \left[
i\left( \mathbf{k}+\mathbf{k}^{\prime }\right) \mathbf{.r}\right] .
\]
The product $K_{EM}$ $\left[ \left| \mathbf{\nabla }\phi \left( \mathbf{r}%
_{0}\right) \right| ^{2}-\phi \left( \mathbf{r}_{0}\right) \mathbf{\nabla }%
^{2}\phi \left( \mathbf{r}_{0}\right) \right] $ is the electromagnetic part
of the additional energy density of interaction between the external
gravitational field and the quantum vacuum (i. e. in addition to the latter
term of eq.$\left( \ref{3a}\right) )$.

The integral giving $K_{EM}$ may be calculated going from the vector
variables $\mathbf{k,k}^{\prime }$ to the new variables $\mathbf{k,s=k}+%
\mathbf{k}^{\prime }.$\textbf{\ } The Jacobian of the transformation is $%
\left| J\right| =1.$ The integral in $r$ may be easily performed and we get 
\[
I=\int r^{2}d^{3}\mathbf{r}\exp \left[ i\left( \mathbf{k}+\mathbf{k}^{\prime
}\right) \mathbf{.r}\right] \rightarrow 4\pi \int_{0}^{\infty }r^{4}dr\frac{%
\sin \left( sr\right) }{sr}\simeq -\frac{48\pi ^{2}}{s^{6}}\delta \left(
s\right) ,
\]
where $\delta \left( {}\right) $ is a Dirac\'{}s delta. Hence after some
algebra we obtain, to the lowest nontrivial order of $s,$ 
\[
K_{EM}\simeq -\frac{16\pi ^{4}}{\left( 8\pi ^{3}\right) ^{2}}\int d^{3}%
\mathbf{k}\int_{0}^{\infty }ds\int_{-1}^{1}du\frac{\left[ 1-u^{2}\right] ^{2}%
}{2k^{3}}\delta \left( s\right) =-\frac{4}{\pi }\int \frac{dk}{k},
\]
where $u=\mathbf{k\cdot s/}ks$. The integral is logaritmically divergent
both at the origin and at infinity. The infrarred divergence is a
consequence of the approximations made. Indeed eq.$\left( \ref{70}\right) $
is not valid for too large values of $r$, whence the integral in $r$ should
not be extended to $\infty $ but only to some distance $R$ of order the
typical range of the gravitational field $\phi \left( r\right) $. Therefore
we should put $R^{-1}$ as lower limit of the $k$ integral. In contrast the
ultraviolet divergence is relevant and we propose to put a cutoff $\Lambda $
in the photon energies. Hence we get 
\begin{equation}
K_{EM}\simeq -\frac{26}{15\pi ^{2}}\int_{R^{-1}}^{\Lambda }\frac{dk}{k}%
\Rightarrow K_{EM}\simeq -\frac{26}{15\pi ^{2}}\ln \left( R\Lambda \right) .
\label{K}
\end{equation}
Actually we should divide by $\sqrt{1+\sum \left| c_{n}\right| ^{2}}$, see
discussion after eq.$\left( \ref{A}\right) .$ We have 
\begin{eqnarray*}
\sum_{n\neq 0}\left| c_{n}\right| ^{2} &=&\frac{1}{12\left( 8\pi ^{3}\right)
^{2}}\int d^{3}\mathbf{k}\int d^{3}\mathbf{k}^{\prime }\frac{(kk^{\prime }+%
\mathbf{k\cdot k}^{\prime })^{2}}{(k+k^{\prime })^{2}kk^{\prime }}\int 
\mathbf{r}^{2}d^{3}\mathbf{r}\exp \left[ i\left( \mathbf{k}+\mathbf{k}%
^{\prime }\right) \mathbf{.r}\right]  \\
&&\times \int d^{3}\mathbf{r}_{0}\left[ \left| \mathbf{\nabla }\phi \left( 
\mathbf{r}_{0}\right) \right| ^{2}-\phi \left( \mathbf{r}_{0}\right) \mathbf{%
\nabla }^{2}\phi \left( \mathbf{r}_{0}\right) \right]  \\
&=&\frac{13}{15\pi ^{2}}R\times 2\int d^{3}\mathbf{r}_{0}\left| \phi \left( 
\mathbf{r}_{0}\right) \right| \mathbf{\nabla }^{2}\phi \left( \mathbf{r}%
_{0}\right) \simeq \frac{26}{15\pi }\times \frac{RMc}{%
\rlap{\protect\rule[1.1ex]{.325em}{.1ex}}h%
}\times 10^{-6},
\end{eqnarray*}
where $M$ is the total mass producing the gravitational field. The second
equality follows from an integration by parts of $\left| \mathbf{\nabla }%
\phi \left( \mathbf{r}_{0}\right) \right| ^{2}$. The latter equality is a
rough estimate substituting a typical value for $\left| \phi \right| \sim
10^{-6},$ that makes the $\mathbf{r}_{0}$ integral trivial taking into
account that $\mathbf{\nabla }^{2}\phi =4\pi \rho $ in our units. In the
final expression I have included explicitly $
\rlap{\protect\rule[1.1ex]{.325em}{.1ex}}h%
$ and $c$ in order to show that $\sum_{n\neq 0}\left| c_{n}\right| ^{2}>>1,$
whence $\left| K\right| $ is actually many orders smaller than eq.$\left( 
\ref{K}\right) .$

The contribution of the free electron-positron field to the vacuum
interacting with a gravitational field, eq.$\left( \ref{n31}\right) $ is
similar to eq.$\left( \ref{n33}\right) .$ We obtain, taking eq.$\left( \ref
{2.6}\right) $ into account, 
\begin{equation}
\hat{\rho}_{vac-grav}^{D}=\left[ \rho _{D0}+\hat{\rho}_{b}\left( \mathbf{r}%
,t\right) +\hat{\rho}_{d}\left( \mathbf{r},t\right) \right] \left[ 1+\phi
\left( \mathbf{r}\right) \right] .  \label{9}
\end{equation}
Getting the Hamiltonian is straightforward via performing the space integral
but I do not write it explicitly. The relevant result is that the state $%
\mid 0\rangle $ with zero electrons and zero positrons is an eigenstate of $%
\hat{\rho}_{vac-grav}^{_{D}}$ because both terms $\hat{\rho}_{b}$\textbf{\ }%
and\textbf{\ }$\hat{\rho}_{d}$ have an annihilation operator on the right.
Hence the energy is 
\[
E_{0}^{D}=\rho _{D}\left[ V+\bar{\phi}\right] ,
\]
where $\rho _{D}$ was given in eq.$\left( \ref{5}\right) .$

Finally we will calculate the contribution of the interaction density eq.$%
\left( \ref{Hint}\right) ,$ that leads to the following Hamiltonian 
\[
\hat{H}_{vac-grav}^{_{int}}=\int d^{3}\mathbf{r}\left[ 1+\phi \left( \mathbf{%
r}\right) \right] \sum_{\mathbf{p,q},\mathbf{k},s,s^{\prime },\varepsilon
}[\zeta _{n}\hat{\alpha}_{\mathbf{k},\mathbf{\varepsilon }}\hat{b}_{\mathbf{p%
}s}\hat{d}_{\mathbf{q}s^{\prime }}\exp \left[ i\left( \mathbf{p+q+k}\right)
\cdot \mathbf{r}\right] +h.c. 
\]
The contribution to the vacuum energy interacting with the gravitational
field is 
\begin{eqnarray}
E_{vac-grav}^{int} &=&-\sum_{n}\frac{\left| \left\langle 0\left| \hat{H}%
_{vac-grav}^{_{int}}\right| n\right\rangle \right| ^{2}}{k+E+E^{\prime }}
\label{10} \\
&=&-\frac{e^{2}}{V^{3}}\sum_{\mathbf{p,q,k}}\frac{\left[ (p^{2}+q^{2})k^{2}+(%
\mathbf{p}\cdot \mathbf{k)(q}\cdot \mathbf{k)}\right] }{2k^{3}EE^{\prime
}\left( k+E+E^{\prime }\right) }  \nonumber \\
&&\times \left| \int_{V}\left[ 1+\phi \left( \mathbf{r}\right) \right] \exp
\left[ i\left( \mathbf{p+q+k}\right) \cdot \mathbf{r}\right] d^{3}\mathbf{r}%
\right| ^{2}.  \nonumber
\end{eqnarray}
After the change of variables eq.$\left( \ref{6a}\right) ,$ taking eq.$%
\left( \ref{70}\right) $ into account we get 
\begin{equation}
E_{vac-grav}^{int}=\rho _{int}\left( V+2\bar{\phi}\right) +B\int d^{3}%
\mathbf{r}_{0}\left\{ \left| \mathbf{\nabla }\phi \left( \mathbf{r}%
_{0}\right) \right| ^{2}-\phi \left( \mathbf{r}_{0}\right) \mathbf{\nabla }%
^{2}\phi \left( \mathbf{r}_{0}\right) \right\} ,  \label{10a}
\end{equation}
where 
\begin{eqnarray}
B &\equiv &\frac{e^{2}}{V^{3}}\sum_{\mathbf{p,q,k}}\frac{\left[
(p^{2}+q^{2})k^{2}+(\mathbf{p}\cdot \mathbf{k)(q}\cdot \mathbf{k)}\right] }{%
2k^{3}EE^{\prime }\left( k+E+E^{\prime }\right) }  \nonumber \\
&&\times \int_{V}r^{2}\exp \left[ i\left( \mathbf{p+q+k}\right) \cdot 
\mathbf{r-}\gamma \left( \left( k+E+E^{\prime }\right) \right) \right] d^{3}%
\mathbf{r,}  \label{B}
\end{eqnarray}
after including the convergence factor eq.$\left( \ref{cutoff}\right) $. I
notice that the right side does not have a minus sign because the latter
term of eq.$\left( \ref{70}\right) $ was changed in going to eq.$\left( \ref
{10a}\right) .$ Now using arguments similar to those of the previous
section, see the paragraph between eqs.$\left( \ref{21.6}\right) $ and $%
\left( \ref{C}\right) ,$ we should redefine the quantity $K_{int}$ as
deriving from 
\begin{equation}
E_{int-grav}=\frac{A\left( 1+2\bar{\phi}/V\right) +B\int d^{3}\mathbf{r}%
_{0}\left[ \left| \mathbf{\nabla }\phi \left( \mathbf{r}_{0}\right) \right|
^{2}-\phi \left( \mathbf{r}_{0}\right) \mathbf{\nabla }^{2}\phi \left( 
\mathbf{r}_{0}\right) \right] }{\sqrt{C\left( 1+2\bar{\phi}/V\right) +D\int
d^{3}\mathbf{r}_{0}\left[ \left| \mathbf{\nabla }\phi \left( \mathbf{r}%
_{0}\right) \right| ^{2}-\phi \left( \mathbf{r}_{0}\right) \mathbf{\nabla }%
^{2}\phi \left( \mathbf{r}_{0}\right) \right] }},  \label{B1}
\end{equation}
rather than eq.$\left( \ref{10a}\right) .$ It may be realized that $B$ and $%
D $ should be of the form (compare with eq.$\left( \ref{C}\right) )$%
\begin{equation}
B=-B_{0}\gamma ^{-2}=dD/d\gamma ,D=B_{0}\gamma ^{-1}.  \label{BD}
\end{equation}
I stress that both terms within the square root in eq.$\left( \ref{B1}%
\right) $ should be positive, being of the form $\sum \left| c_{n}\right|
^{2}$ each. Therefore $D>0$ whence $B<0,B_{0}>0.$

The interpretation of eq.$\left( \ref{B1}\right) $ is more clear if it is
approximated by an expression quadratic in $\phi ,$ taking into account that 
$A,B,C,$and $D$, eqs.$\left( \ref{A}\right) ,\left( \ref{C}\right) $ and $%
\left( \ref{BD}\right) ,$ fulfil $\left| B\right| <<\left| A\right|
\Rightarrow \left| D\right| <<\left| C\right| $ and that $\left| \bar{\phi}%
\right| <<V$. Thus we get 
\begin{eqnarray*}
E_{int-grav} &\simeq &\frac{A}{\sqrt{C}}\left( 1+\bar{\phi}/V\right) +\left[ 
\frac{B}{\sqrt{C}}-\frac{AD}{2C\sqrt{C}}\right] \int d^{3}\mathbf{r}%
_{0}\left[ \left| \mathbf{\nabla }\phi \left( \mathbf{r}_{0}\right) \right|
^{2}-\phi \left( \mathbf{r}_{0}\right) \mathbf{\nabla }^{2}\phi \left( 
\mathbf{r}_{0}\right) \right] \\
&\Rightarrow &\rho _{int-grav}\simeq \rho _{int}\left( 1+\bar{\phi}/V\right)
+\frac{B_{0}}{2\sqrt{\left| A_{0}\right| }}\gamma ^{-1/2}\left[ \left| 
\mathbf{\nabla }\phi \left( \mathbf{r}_{0}\right) \right| ^{2}-\phi \left( 
\mathbf{r}_{0}\right) \mathbf{\nabla }^{2}\phi \left( \mathbf{r}_{0}\right)
\right] ,
\end{eqnarray*}
where $B_{0}/\sqrt{\left| A_{0}\right| }$ is a dimensionless quantity. I
stress that the latter term is positive.

In summary we may write the QED vacuum energy in the presence of a
gravitational potential $\phi \left( \mathbf{r}\right) $ in the form 
\begin{eqnarray}
E_{grav-vac}^{QED} &=&\rho _{QED}V+\rho _{QED}\bar{\phi}+VK_{QED}\left[
\left| \mathbf{\nabla }\phi \left( \mathbf{r}_{0}\right) \right| ^{2}-\phi
\left( \mathbf{r}_{0}\right) \mathbf{\nabla }^{2}\phi \left( \mathbf{r}%
_{0}\right) \right] ,  \nonumber \\
&&\rho _{QED}=\rho _{EM}+\rho _{D}+\rho _{int}  \label{12}
\end{eqnarray}
where 
\begin{equation}
K_{QED}=K_{EM}+K_{int},K_{int}\equiv \frac{B_{0}}{2\sqrt{\left| A_{0}\right| 
}}\gamma ^{-1/2}.  \label{12K}
\end{equation}
The first term of eq.$\left( \ref{12}\right) $ is the standard QED vacuum
energy, i. e. in the absence of external gravitational field, and the second
term is the interaction of that field with the unperturbed vacuum. I propose
that the latter term is an alternative to ``dark matter'', see below. The
dimensionless quantity $K_{QED}$ consists of a pure electromagnetic
contribution $K_{EM}$ defined in eq.$\left( \ref{K}\right) $ (negative and
divergent like $\log \Lambda $), associated to the production of virtual
photon pairs with opposite momenta, plus the contribution $K_{int}$
(positive and divergent like $\sqrt{\Lambda }$) corresponding of the
production of virtual triples photon-electron-positron with total zero
momentum. As a conclusion of our calculation we see that the relevant part
of the interaction energy density between the vacuum fields and an external
gravitational field (i. e. the latter term of eq.$\left( \ref{12}\right) )$
is positive.

Eq.$\left( \ref{12}\right) $ is the correct substitute for the wrong eq.$%
\left( \ref{n10}\right) .$ The difference between them is the latter term of
eq.$\left( \ref{12}\right) .$ If we were studying the whole set of
elementary particles and their interactions, I have assumed that the quantum
vacuum energy is zero (or it is cancelled by some unknown mechanism), whence
the generalization of eq.$\left( \ref{12}\right) $ would consist only of
terms like the latter one. That is the energy density at point $\mathbf{r}%
_{0}$ due to the interaction of the quantum vacuum with an external
gravitational potential $\phi \left( \mathbf{r}_{0}\right) $ would be
approximately 
\begin{equation}
\rho _{vac-grav}\simeq K\left[ \left| \mathbf{\nabla }\phi \left( \mathbf{r}%
_{0}\right) \right| ^{2}-\phi \left( \mathbf{r}_{0}\right) \mathbf{\nabla }%
^{2}\phi \left( \mathbf{r}_{0}\right) \right] ,  \label{n.43}
\end{equation}
$K$ being a new constant that would generalize the parameter $K_{QED}$, eq.$%
\left( \ref{12}\right) ,$ for all quantum fields and interactions. This is
valid only for weak external gravitational fields where the Newtonian
approximation applies, although the divergences put a problem not to be
discussed further here. In any case the divergences found are logaritmic, in
eq.$\left( \ref{K}\right) $, or of order $\sqrt{\Lambda }$ in eq.$\left( \ref
{12K}\right) ,$ much weaker than the quartic divergence of the vacuum
energy, see eqs.$\left( \ref{0}\right) $ and $\left( \ref{5}\right) .$

I propose that this interaction might the origin of the socalled ``dark
matter'', thus making unnecessary to postulate the existence of new
particles not appearing in the standard model. In order to support that
conjecture, in the next section I will exhibit a simple model inspired on eq.%
$\left( \ref{n.43}\right) $.\textrm{\ }

\section{Simple model alternative to dark matter}

\subsection{The problem of the flat rotation curves in galactic
haloes\smallskip}

The existence of dark matter of unknown origin is the current solution
proposed for two different problems\cite{Einasto},\cite{Okali} which may be
called astrophysical and cosmological, respectively. The astrophysical
problem is the more or less flat rotation curves of stars or gas in the halo
of galaxies and culsters of galaxies. The cosmological one is the need of a
large amount of non-baryonic cold matter in the universe, at the
recombination time, in order to explain the formation of structures. The
amount of non-baryonic matter is fairly the same in both cases, which has
lead to the popular assumption that both may be solved with the same
hypothesis, that is dark matter. The standard $\Lambda CDM$ cosmological
model predicts that about 70\% of mass-energy budget of the Universe is
composed of dark energy, about 30\% is the total mass of the Universe which
is dominated with more than 5/6 by dark matter, while ordinary baryonic
matter constitutes only less than1/6 of the total mass\cite{Planck}. In
order to account for the flat rotation curves of spiral galaxies, dark
matter should dominate the total mass of the galaxy and should be
concentrated in the outer baryonic regions of galactic disks, as well as in
the surrounding haloes.

The fact that none of the known particles is a good candidate to form the
dark matter and the failure to discover new particles with the required
properties, in spite of the big effort made at the observational level, has
lead to alternatives consisting of a modification of current gravity
theories (i. e. general relativity or Newtonian gravity) or dynamics. The
most elaborate of these is the ``modified Newtonian dynamics'' (MOND)\cite
{Milgrom}, but there are also proposals resting upon $f(R)$ gravity or its
generalizations\cite{Faraoni}, \cite{Capozziello6},\cite{Capozziello17}.
Actually the proposed modifications of gravity or dynamics attempt mainly to
solve the astrophyssical problem, with less implication in the cosmological
one.

MOND\cite{Milgrom} proposes, for galaxies approximately spherical, that the
flat (i. e. independent of the radius) rotation velocities, $v$, in galactic
haloes is roughly given by 
\begin{equation}
v^{2}\simeq \sqrt{a_{0}GM_{B}},  \label{1d}
\end{equation}
where $G$\ is Newton constant, $M_{B}$\ the baryonic mass of the galaxy, and 
$a_{0}$\ a new universal constant given by 
\[
a_{0}=2\times 10^{-10}\text{ m/s}^{2}. 
\]
The dependence on the square root of the total baryon mass of the galaxy
fits in the empirical baryonic Tully-Fisher law\cite{Milgrom}, that is $%
v^{4}\varpropto M_{B}$. Also MOND is able to successfully explain the
dynamics of galaxies outside clusters and a discovered tight relation
between the radial acceleration inferred from their observed rotation curves
and the acceleration due to the baryonic components of their disks\cite
{McGaugh}. However the MOND predictions do not fit too well in some
observations\cite{Petersen}.

Another proposed alternative to dark matter is extended general relativity
(see \cite{Capozziello11} for a review). In these theories a different
scalar obtained from the Riemman tensor is substituted for the Ricci scalar,
R, in the Hilbert-Einstein action. The most interesting case consists of
choosing a function of R, leading to the socalled f(R) theories. Then it is
possible to demonstrate that the existence of a Noether symmetry in f(R)
theories of gravity gives rise to a further gravitational radius, besides
the standard Schwarzschild one, determining the dynamics at galactic scales%
\cite{Capozziello17}. This radius plays an analog role, in the case of weak
gravitational field at galactic scales, like the Schwarzschild radius in the
case of strong gravitational field in the vicinity of compact massive
objects. Such a feature emerges from symmetries that exist for any power-law 
$f(R)$ function. In particular, for $f(R)\varpropto R^{3/2}$, the MOND
acceleration regime is recovered. Using this new gravitational radius, $f(R)$
theories of gravity are able to explain the baryonic Tully-Fisher relation
of gas-rich galaxies in a natural way.

In this paper I propose another hypothesis, namely that the interaction of
gravitational fields with the vacuum quantum fields might explain the
rotation curves, but the cosmological problem is not studied. The idea of
explaining the flat rotation curves of galaxies as a quantum vacuum effect
is not new. For instance a model has been proposed in order to explain eq.$%
\left( \ref{1d}\right) $ under the assumption of a gravitational repulsion
between particles and antiparticles which, via vacuum fluctuations, would
give rise to a gravitationally polarizable vacuum\cite{Hajdukovic}. The
author attributes the repulsion to a negative gravitational mass (but
positive inertial mass) of antiparticles.

\subsection{Our model}

I propose a model for stationary systems whose fundamental assumption is the
following equation (compare with eq.$\left( \ref{n.43}\right) )$%
\begin{equation}
\rho _{g}\left( \mathbf{r}\right) =K\left[ \left| \mathbf{\nabla }\phi
\left( \mathbf{r}\right) \right| ^{2}+a\mathbf{\nabla }^{2}\phi \left( 
\mathbf{r}\right) \right] ,  \label{3d}
\end{equation}
where $\rho _{g}$ is the net energy density of interaction between the
quantum vacuum and a gravitational field with Newtonian potential $\phi
\left( \mathbf{r}\right) .$ In this model $K$ and $a$ are positive
parameters and I will use units $c=4\pi G=1$ in this section (Planck
constant $
\rlap{\protect\rule[1.1ex]{.325em}{.1ex}}h%
$ does not enter in the model although it would enter in the calculation of $%
K$ ). The model is suggested for the interaction between a weak external
gravitational field and the QED vacuum, calculated in section 2 leading to
eq.$\left( \ref{n.43}\right) .$ In comparison with that equation I
substitute $-a$ for the Newtonian potential, $\phi \left( \mathbf{r}\right) $%
, which produces a big simplification of the model. It is justified by the
fact that the potencial changes more slowly than the field, as will be
checked below. The field, $-\mathbf{\nabla }\phi \left( \mathbf{r}\right) ,$
derives from all matter in space that we assume to consists of baryonic, $%
\rho _{B}\left( \mathbf{r}\right) ,$ plus the grav-vac interaction $\rho
_{g}\left( \mathbf{r}\right) $ itself. Consequently I include in the model
another equation giving the relation between field and mass-energy density,
that is 
\begin{equation}
\mathbf{\nabla }^{2}\phi \left( \mathbf{r}\right) =\rho _{B}\left( \mathbf{r}%
\right) +\rho _{g}\left( \mathbf{r}\right) .  \label{4d}
\end{equation}
From eqs.$\left( \ref{3d}\right) $ and $\left( \ref{4d}\right) $ we get 
\begin{equation}
\left( 1-aK\right) \mathbf{\nabla }^{2}\phi \left( \mathbf{r}\right) =\rho
_{B}\left( \mathbf{r}\right) +K\left| \mathbf{\nabla }\phi \left( \mathbf{r}%
\right) \right| ^{2}.  \label{5d}
\end{equation}
This is the fundamental equation of the model.

For a spherically symmetric system (galaxy or cluster) eq.$\left( \ref{5d}%
\right) $ becomes 
\begin{equation}
\frac{d\phi ^{\prime }}{dr}+2\frac{\phi ^{\prime }}{r}=\left( 1-aK\right)
^{-1}\left[ \rho _{B}\left( \mathbf{r}\right) +K\phi ^{\prime 2}\right] .
\label{6d}
\end{equation}
where $-\phi ^{\prime }$ is the gravitational field. This equation has no
regular solution (both at the origin and at infinity) if $\rho _{B}\left( 
\mathbf{r}\right) =0$ everywhere. This is satisfactory because it shows that
``dark matter'' appears only near regions with baryonic matter, in agreement
with observations. In the external region, $r>R,$ where $\rho _{B}\left( 
\mathbf{r}\right) =0$ there is a solution regular at infinity, namely 
\begin{equation}
\phi ^{\prime }\left( r\right) =\frac{1-aK}{Kr}.  \label{7d}
\end{equation}
For this field the rotation curves are flat in agreement with observations.
From eq.$\left( \ref{7d}\right) $ the rotation velocity $v$ fulfils 
\begin{equation}
v^{2}=\frac{1-aK}{K}c^{2}\text{.}  \label{8d}
\end{equation}
Thus we fix the parameters so that it fits the observed rotation velocity in
the halo of a typical galaxy, say $v\sim 10^{-3}c,$ whence we get 
\begin{equation}
\frac{1-aK}{K}\sim 10^{-6}.  \label{10d}
\end{equation}

Taking eqs.$\left( \ref{5d}\right) $ and $\left( \ref{7d}\right) $ into
account the dark mass density in the region without baryonic matter becomes,
for any spherical body, 
\begin{equation}
\rho _{g}\left( \mathbf{r}\right) =\frac{1-aK}{Kr^{2}}.  \label{9d}
\end{equation}
The gravitational potential is given by the integral of eq.$\left( \ref{7d}%
\right) ,$ that is 
\[
\phi \left( r\right) =-\phi _{0}+\frac{\left( 1-aK\right) }{K}\ln \left(
r/r_{0}\right) , 
\]
where the parameters $r_{0}$ and $\phi _{0}$ depend on the particular
galaxy. The logaritmic dependence of $\phi \left( r\right) $ shows that the
potential is slowly varying in the halo of the galaxy, as we commented after
eq.$\left( \ref{3d}\right) .$

For a galaxy with baryonic radius $R\sim 10^{22}m$ the dark mass density in
the halo would be, taking eqs.$\left( \ref{10d}\right) $ and $\left( \ref{9d}%
\right) $ into account, 
\[
\rho _{g}\sim 10^{-24}kg/m^{3}. 
\]
That is of\ order the mean density of the galaxy if $M\sim 10^{42}kg$ is the
mass. The density $\rho _{g}$\ predicted by the model near individual stars
is too big. For instance in the surface of the sun $(M=2\cdot
10^{30}kg,R=7\cdot 10^{8}m)$\ it gives $\rho _{g}\sim 3000kg/m^{3}.$
Obviously the model is not valid for gravitational fields much stronger then
those of galaxy haloes.

Near the center of a spherical body the gravitational field, $-\mathbf{%
\nabla }\phi \left( \mathbf{r}\right) ,$ is zero or small whence eq.$\left( 
\ref{5d}\right) $ leads to 
\[
\rho _{g}\sim \frac{aK}{1-aK}\rho _{B}=10^{6}a\rho _{B}\sim 10^{6}\left|
\phi \left( 0\right) \right| \rho _{B}\left( 0\right) . 
\]
where I\ have taken into account eq.$\left( \ref{10d}\right) $ and the fact
that $a$ is our model substitute for the potential $\left| \phi \left(
r\right) \right| $ (compare eqs.$\left( \ref{n.43}\right) $ and $\left( \ref
{3d}\right) $). Therefore the model predicts a relatively high dark density
at the center, a phenomenon observed and usually known as a ``cusp''. Indeed
if $\phi \left( 0\right) $ is larger than $10^{-6}$ in our units, then $\rho
_{g}$ is larger than $\rho _{B}$ at the center of the galaxy.

As we have seen there are several predictions of the model that fit fairly
well in the observations. In fact, it leads to dark mass distributions that
go well beyond the radius of typical galaxies. It predicts flat rotation
curves in the haloes and a maximum of the dark density at the center of the
galaxy (or cluster). It is also a good feature that the fundamental
equations do not have any physically sound solution in the absence of
baryonic matter. Nevertheless there are difficulties that point towards the
need of some modification of the model, the three most relevant being the
following.

1. The model predicts a large amount of dark mass around bodies like the sun
that is not observed. Indeed the ratio $\rho _{g}/\bar{\rho}_{B}$ predicted
by the model is almost independent of the mean baryonic density $\bar{\rho}%
_{B}.$ A modification of the model eq.$\left( \ref{3d}\right) $ of the form 
\[
\rho _{g}\left( \mathbf{r}\right) =K\left[ \left| \mathbf{\nabla }\phi
\left( \mathbf{r}\right) \right| ^{2}+a\mathbf{\nabla }^{2}\phi \left( 
\mathbf{r}\right) \right] \times f\left( \mathbf{\nabla }^{2}\phi \left( 
\mathbf{r}\right) \right) , 
\]
with $f(x)$ a decreasing function (i. e. $df/dx<0$) , might solve this
difficulty. The decrease should be very slow at typical galaxy densities
thus not spoiling the predicted flat rotation curves. This is possible
because the ratio of the Sun density to the typical density in a galaxy is
about $10^{27}.$

2. The rotation velocity does not depend on the baryonic mass, $M_{B}$, of
the associated galaxy, contradicting the baryonic Tully-Fisher empirical
law, that is a velocity proportional to $M_{B}^{1/4}.$

3. The total dark mass is divergent because the predicted dark density
behaves as $r^{-2}$ at infinity, so that $\int \rho _{g}d^{3}\mathbf{r}$
diverges.

\section{Conclusions}

The change of the vacuum energy in presence of a gravitational field is
calculated in the particular example of QED vacuum, the result suggesting
that there is a finite interaction of the field with the quantum vacuum.
There should be many contributions to that interaction, but the net result
is an energy density that, if it turns out to be positive, we may interpret
as the origin of the flat rotation curves in galactic haloes, commonly
attributed to some ``dark matter''. A model is proposed that fits in the
main properties of the rotation curves, although it presents difficulties.


\begin{thebibliography}{99}
\bibitem{Milonni}  P. Milonni, \textit{The quantum vacuum. An introduction
to quantum electrodynamics}, Academic Press, San Diego, 1994.

\bibitem{Weinberg}  S. Weinberg, \textit{Rev. Mod. Phys.} \textbf{61}, 1
(1989).

\bibitem{Zeldovich}  B. Y. Zeldovich, \textit{Sov. Phys. Usp.} \textbf{24},
216 (1981).

\bibitem{Santos}  E. Santos, \textit{Astrophys. Space Sci}. \textbf{332},
423-435 (2011).

\bibitem{expansion}  P. J. E. Peebles y Bharat Ratra, \textit{Rev. Mod. Phys.%
} \textbf{75}, 559-606 (2003).

\bibitem{Birrell}  N. D. Birrell and P. C. W. Davies, \textit{Quantum Fields
In Curved Space}, Cambridge University Press, 1982.

\bibitem{Jacobson}  T. Jacobson, Introduction to quantum fields in curved
spacetime and the Hawking effect, arXiv:gr-qc/0308048v3 (2003).

\bibitem{FOS}  E. Santos, \textit{Foundations of Science}, \textbf{20},
357-386 (2015).

\bibitem{Einasto}  J. Einasto, Dark Matter, arXiv: 0901.0632v2 (2010).

\bibitem{Okali}  C. Okali, Dark matter hole concentrations: a short review,
arXiv: 1711.05277 (2017).

\bibitem{Planck}  Planck Collaboration, P. A. R. Ade, N. Aghanim, et al., 
\textit{A\&A}, \textbf{594}, A13 (2016).

\bibitem{Milgrom}  M. Milgrom, \textit{Astrophys. J. }\textbf{270}, 365
(1983)\textit{; }The\textit{\ }MOND paradigm, arXiv: 0801.3133v2 (2008).

\bibitem{Faraoni}  T. P. Sotiriou and V. Faraoni, \textit{Rev. Mod. Phys. }%
\textbf{82}, 451 (2010).

\bibitem{Capozziello6}  S. Capozziello, V. F. Cardone, A. Troisi,\textit{\
J. Cosmol. Astropart. Phys.} \textbf{0608}, 001 (2006).

\bibitem{Capozziello17}  S. Capozziello, P. Jovanovi\'{c}, V. Borka
Jovanovi\'{c}, D. Borka, \textit{\ J. Cosmol. Astropart. Phys.} \textbf{1706}%
, 044 (2017).

\bibitem{McGaugh}  McGaugh, S. S., Lelli, F., \& Schombert, J. M., \textit{%
Phys. Rev. Lett.} \textbf{117}, 201101 (2016).

\bibitem{Petersen}  J. Petersen and M. Frandsen, Discriminating between dark
matter and MOND via galactic rotation curves, arXiv: 1710.03096 (2017).

\bibitem{Capozziello11}  S. Capozziello and M. De Laurentis, \textit{Phys.
Rep.} \textbf{509}, 167 (2011).

\bibitem{Hajdukovic}  D. S. Hajdukovic, \textit{Astrophys. Space Sci}. 
\textbf{334}, 215-218 (2011).
\end{thebibliography}
\end{document}